\documentclass[12pt]{article}
\def\be{\begin{equation}} \def\ee{\end{equation}}
\def\bi{\begin{itemize}} \def\ei{\end{itemize}}
\def\bea{\begin{eqnarray}} \def\eea{\end{eqnarray}} \def\ba{\begin{array}}
\def\ea{\end{array}} \def\ben{\begin{enumerate}} \def\een{\end{enumerate}}

\newcommand{\eqn}[1]{(\ref{#1})}

\newcommand{\hepth}[1]{{\tt [arXiv:{#1} [hep-th]]}}

\def\m{\mu}

\def\n{\nu}

\def\br{\nonumber\\}

\def\G{\Gamma}
\def\tr{{\rm Tr}}

\begin{document}
{}~
\hfill\vbox{\hbox{hep-th/yymm.nnnn} \hbox{\today}}\break

\vskip 2.5cm
\centerline{\large \bf
Super-Yang-Mills and M5-branes}
\vskip .5cm

\vspace*{.5cm}

\centerline{  
Harvendra Singh
}
\vspace*{.25cm}
\centerline{ \it  Theory Division, Saha Institute of Nuclear Physics} 
\centerline{ \it  1/AF Bidhannagar, Kolkata 700064, India}
\vspace*{.25cm}
\vspace*{.25cm}

\vspace*{.5cm}

\vskip.5cm

\vskip1.5cm

\centerline{\bf Abstract} \bigskip

We  uplift  5-dimensional super-Yang-Mills theory to a 6-dimensional 
gauge theory  with the help of a   space-like constant
vector $\eta^M$, whose norm determines the YM coupling constant. 
After the localization of $\eta^M$ the
6D gauge theory acquires Lorentzian invariance 
as well as scale invariance. 
We discuss KK states, instantons and the flux quantization.
The  theory admits extended solutions like $1/2$ BPS  `strings' and 
monopoles.

\vfill 
\eject

\baselineskip=16.2pt


\section{Introduction}
Recent progress in formulating the holographic
 super-membrane theories \cite{blg}-\cite{hs}, 
especially BLG tri-algebra based theories \cite{blg} 
and the ABJM  Chern-Simons matter theories \cite{abjm}, 
has led to a renewed interest in understanding the
mysteries behind so far unknown
  M5-brane theory
\cite{Lambert:2010wm}
\cite{Douglas:2010iu}
\cite{lps}. 
The current understanding is that the 
dynamics of a single M5-brane is governed by an Abelian
 (2,0) super-conformal tensor theory
having  maximal supersymmetry in six-dimensions. The antisymmetric
2-rank tensor fields are natural objects to occur 
in six dimensions \cite{howe1}. There is a reason to this;
when a  fundamental M2-brane ends on 
 M5-brane the intersection produces a line defect on 6D world-volume of the
M5-brane. These defects  entirely live on the M5-brane world-volume
and  behave like extended `strings'. 
The belief is that these strings  constitute the 
fundamental excitations on the M5-branes. 
The strings would naturally couple to 
2-rank tensor field, $B_{\m\n}$, whose field strength is a 3-form, 
$H_{(3)}=dB_{(2)}$. 
There are  already  known  self-dual string like  solutions 
 on  M5-brane \cite{howe}. The  tensor field, $B$,
{\it five} scalars, $X^I$, and  a
spinor, $\Psi$, constitute what is known as
 (2,0) tensor super-multiplet in 6-dimensions \cite{howe1}. 
The dynamical equations 
of this Abelian tensor theory are very simple and are given by
\bea
\label{self1a}
H_{(3)}\equiv dB_{(2)}=\star_6 H_{(3)}, ~~~
\partial_M\partial^M X^I=0=\not\!\partial\Psi
\eea
where $\star_6$ is the Hodge-dual operation in six dimensions, $\Psi$ is 
the Majorana spinor. 
The massless scalars $X^I$ correspond to  five translational (Goldstone)
modes, which an extended M5-brane acquires 
when placed in a flat 11-dimensional spacetime.
The (2,0) tensor theory   
is  finite  and superconformal,  but  the theory is trivial as it is
 noninteracting. However, it is  being currently argued that
all the states of a non-Abelian (2,0) tensor theory, once compactified
on a circle, are probably contained in the 5-dimensional 
super-Yang-Mills (SYM) theory. Note, as such the 5D SYM theory is known to be
powercounting nonrenormalizable and is  strongly coupled in the UV region. 
But if 5D SYM indeed contains all the states of a 6D CFT
without requiring any new UV degrees of freedom, then 
it should also be a finite theory by itself \cite{lps}. Although intuitive, 
but it is very difficult to directly check  the finiteness of the 5D SYM. 
On the other hand
not much is known about the non-Abelian construction of 
(2,0) tensor theory itself, which is supposed to  
describe an interacting theory living on a stack of multiple M5-branes. 
In analogy with YM fields, the tensor fields should  be
self-interacting just like Yang-Mills fields do. The non-Abelian 
6D CFT should simply 
possess a $SU(N)$  gauge symmetry along with  $SO(5)$ R-symmetry. These 
are some of the simple requirements which have so far eluded us in 
a meaningful construction of the M5-brane theory.  
\footnote{ See some of the earlier developments in this field 
in the references 
\cite{witt,ps,pst96,tonin,howe1,howe2,Howe:1997fb,Lee:2000kc}.}

Our goal in this paper is rather different. We would like 
to construct a non-Abelian  theory in 
six  dimensions which could describe M5-branes and it need not be 
a tensor like theory. 
We shall introduce an overall constant space-like vector $\eta^M$ whose 
norm will determine the
Yang-Mills coupling constant. We then show that with the help of this 
vector the 5D 
Yang-Mills theory can be embedded into a 6D framework in a  Lorentz 
covariant manner. After localization of $\eta^M$ the
 theory  recovers Lorentz  invariance as well as scale invariance.
The theory also carries correct counting of physical degrees of freedom
such that we have  maximal supersymmetry.
The 6D non-Abelian gauge theory upon dimensional reduction gives rise to
the usual 5D SYM. 

The paper is organized  as follows. In section-II 
we review the basics of $SU(N)$ super-Yang-Mills theory in five
dimensions. In  section-III 
we first introduce  6D Abelian gauge theory and discuss its solutions in detail. 
We  find that the Abelian  theory admits ${1\over 2}$-BPS  `strings' and 
monopole solutions but there are no stable point like solutions. 
We then introduce a  
non-Abelian gauge theory with the help of  space-like constant 6-vector.
We also present the supersymmetry transformations and discuss 
straight forward dimensional reduction to 5D. In section-IV we restore the 
Lorentz invariance and replace the constant vector with
an auxiliary Abelian field. It requires an introduction of a 
Lagrange multiplier 4-form potential. With this 
modification the 6D gauge theory is found to be scale invariant. 
We also discuss KK states, 
 instantons and the quantization of the flux. The conclusion is given 
in section-V.

\section{5D super Yang-Mills theory}
Maximally supersymmetric Yang-Mills  theories 
in $(p+1)$ dimensions are well known to describe
low energy dynamics on the stacks of multiple D$p$-branes. 
For D4-branes  the SYM action 
in five dimensions is given by
\bea\label{eq1}
S_{YM}&=&\int  
d^5x\, \tr \bigg[  -{1\over 4g^2} F_{\m\n}F^{\m\n} 
-{1\over2}
 D_\mu X^I D^\mu X^I +{g^2\over4}([X^I,X^J])^2
\br &&+{i\over 2}\bar\Psi\Gamma^\mu D_\mu\Psi
-{g\over 2}\bar\Psi\Gamma^5\Gamma^I[X^I,\Psi]
  \bigg]
\eea
where $A_\mu,~(\mu=0,1,\cdots,4)$, is the gauge field and 
 $X^I,~(I=6,7,8,9,10)$, are five scalars and $\Psi$ are the fermions. 
All the fields 
are in the $N\times N$  adjoint representation of the  gauge group $SU(N)$.
The commutators belong to the Lie algebra.
The gauge field strength can be written as
 $$F_{\mu\nu}=\partial_{[\mu}A_{\nu]}-i [A_\mu,A_\nu]$$  
and the covariant derivative is
\be D_\mu X^I=\partial_\mu X^I -i[A_\mu,X^I].\ee

The  supersymmetry variations  are given by 
\bea\label{susy1}
&& \delta_s X^I= i\bar\epsilon \Gamma^I\Psi\br
&& \delta_s A_\mu= i g \bar\epsilon \Gamma_\mu\Gamma_5\Psi\br
&& \delta_s \Psi= {1\over2g} F_{\mu\nu}\Gamma^{\mu\nu}\Gamma_5\epsilon 
+D_\mu X^I\Gamma^\mu \Gamma^{I}\epsilon
-{i\over 2}g[X^I,X^J]\Gamma^{IJ}\G^5\epsilon
\eea
under which the  action \eqn{eq1} closes on-shell. All spinors  have 
32 real components. The  constant  spinors in supersymmetry transformations
satisfy the projection condition
$\Gamma_{012345}\epsilon=\epsilon$ while spinor 
$\Psi$ has opposite chirality,
$\Gamma_{012345}\Psi=-\Psi$. Most of our notations match with \cite{lps}.

The 5D super Yang-Mills theory is 
known to be powercounting nonrenormalizable 
as its coupling  constant $g$ is  dimensionful with mass
dimension $-{1\over2}$. We know
  that in the strong coupling regime the
string (bulk) theory becomes effectively 11-dimensional M-theory,
 hence from the AdS/CFT holography
the boundary CFT  should  run to a strongly coupled
conformal fixed point (in the UV regime)
where the theory should rather be described by 
an stack of M5-branes 
instead of D4-branes. That is, an extra spatial
world-volume direction should open up in 5D SYM theory 
and consequently it should become a 6-dimensional gauge/tensor theory. 
The precise mechanism how this would happen is not quite known yet. 
There are arguments which suggest that  the
 natural objects which  describe  M5-branes  are the 2-rank 
 tensor fields instead of the gauge fields. Actually there are 
known   string-like self-dual
solitonic configurations which can live on M5-branes \cite{howe}.
We shall discuss about these solutions in the next section.
Note that the physical degrees of freedom contributed by
a self-dual tensor fields in 6D  add up to {\it three} only. 
We also know that these are precisely the d.o.f. 
also contributed by the Yang-Mills field in 5D. These d.o.f counts 
and the maximal supersymmetry
are some of the evidences that the 5D SYM could perhaps be lifted 
to 6D gauge or a tensor-like theories. 
 Thus if we could embed the
5D SYM in some suitable 6-dimensional framework we would be 
somewhat successful. We mention that writing down an explicit version of an 
interacting non-Abelian tensor theory describing multiple M5-branes 
has remained an illusive goal so far. Although along these directions 
 there have been some  interesting 
attempts recently, e.g. introducing tri-Lie 
algebras 
in the M5-brane tensor theories \cite{Lambert:2010wm,hhm}.

\section{6D non-Abelian gauge theories}
As stated above our modest aim is to uplift the cosmetic structure of 
the 5D SYM theory 
to six dimensions such that the theory looks like a
gauge theory
but gives 5D super-Yang-Mills upon dimensional reduction. 
In order to achieve this we first introduce a spacelike 
(coupling) constant  vector $\eta^M$ in six dimensions,
which is normalized as 
\be\label{vec1}
\eta^M\eta_M=g^2 >0
\ee 
where $g$ is a constant and will be
related to  5D  Yang-Mills coupling constant.
Though out in this text $M,N=0,1,\ldots,5$ will represent 
6-dimensional Lorentzian indices. 

We have to understand first whether we need 
to work with vector fields in order to describe M5 branes. Factually,  
there are no apparent  
dynamical processes involving M5-branes to which we could assign the 
presence of vector fields. But, interestingly, we do include vector 
fields to work with  Chern-Simons matter theories, ABJM or BLG,
which are super-membrane theories \cite{abjm,blg}.  
Under the same spirit, for M5-branes case here, although
 we do not apriori know what these 6D gauge
fields may represent, but the idea is that the gauge fields
on their own may also act like tensor fields and vice-versa.

\subsection{Abelian gauge theory in 6D}
First we discuss a simple example of an Abelian theory in 6D.
It is rather straight forward to 
write down a 6-dimensional supersymmetric gauge action
with the help of a fixed space-like 
 vector $\eta^M$ given in eq.\eqn{vec1}. 
Correspondingly 
a covariant 6-dimensional   action  involving Abelian vector field is
\bea\label{act3aap}
S[A]&\equiv&\int d^6x  \bigg[
-{1\over 2.3! \eta^4} ( \eta_{[M}F_{NP]})^2  
 -{1\over2} (\partial_M X^I)^2 +{i\over 2}\bar\Psi\Gamma^M\partial_M\Psi
  \bigg]
\eea
where 
 \be
\eta_{[M}F_{NP]}=\eta_{M}F_{NP}+ ~{\rm cyclic~ permutations~ of~ indices}
\ee
and the Abelian field strength is
 $F_{NP}=\partial_N A_P-\partial_P A_N$.
Here  $X^I~(I=6,7,...,10)$ are   five real scalars.
 These would correspond to the fact that M5-brane has
five coordinates transverse to its world-volume.
The bosonic equations of motion are
\bea\label{hl1}
&& \partial_M \partial^M X^I=0=\not\!\partial\Psi\br
&& 
\partial_M \eta^{[M}F^{NP]}=0
\eea
while the Bianchi identity is $ dF=0$. Thus equations of motion
are  covariant.
But these equations are very different from those of 
Abelian tensor theory 
in \eqn{self1a} which involves a self-dual 3-form field strength.
The  supersymmetry variations can be written as
\bea\label{susy2ap}
&& \delta_s X^I= i\bar\epsilon \Gamma^I\Psi\br
&& 
 \delta_s A_{M}= i\bar\epsilon \eta^N\Gamma_{MN}\Psi\br
&& \delta_s \Psi= {1\over3!\eta^2} \eta_{[M}F_{NP]}\Gamma^{MNP}\epsilon 
+\not\!\partial X^I \Gamma^{I}\epsilon
\eea
under which the action \eqn{act3aap} closes on-shell. 
Note that  $\epsilon$ and $\Psi$ 
are spinors of $SO(1,10)$ and
have opposite chiralities. 

We notice that the gauge kinetic term in \eqn{act3aap} 
is rather unusual. This  axial 
form  of  gauge  action helps us in maintaining the 
right field content in the theory. It can be seen as follows. The 6D
gauge field has six off-shell degrees of freedom.
Any preferential choice of  vector $\eta^M$ will
 break the covariance spontaneously.
For example, if we locally choose  
 $\eta^M=({\bf 0},g)$, implying that the vector $\eta^M$
is aligned  along $x^5$,  it would turn the 5-th component
of the gauge field, $A_5$,  essentially nondynamical and  
auxiliary. That is there would be no kinetic term involving $A_5$ 
in the action.
The residual gauge symmetry of the theory fixes 2 more d.o.f..
Thus the actual on-shell gauge d.o.f.  remain $(6-3)=3$ only. 
These 3 gauge degrees of freedom and the 5 scalars,  $X^I$, 
constitute in total 8 bosonic degrees of freedom. We recall  that 
$\Psi$  is a Majorana spinor, and it is chiral in nature, so it  
  also contributes 8  fermionic d.o.f.. 
Thus our choice of covariant (axial) gauge field strength guarantees
 that the 6D Abelian theory  \eqn{act3aap} 
has the right physical (on-shell) content required for the maximal
supersymmetry. (Had we taken a Maxwell type action $(F_{MN})^2$, 
it wouldn't have helped us, because 
the gauge field then would have contributed 4 physical d.o.f., 
one more than what is necessary.) 
The axial gauge action is however fully gauge invariant. 
(See also the discussion in the Appendix.) Obviously the 
Lorentz invariance is compromised here. In section-4 we will show 
how $\eta^M$ can  be promoted to the status of 
a local $U(1)$ field where Lorentz invariance  
is ultimately regained. 

We now list some of the static vacua of the Abelian theory in \eqn{act3aap}.
\bi
\item i) We first write down the vacua which is an extended solitonic
configuration, describing an  M2-brane ending on  M5-brane. (In the language 
of self-dual tensor theories \cite{howe}, such a configuration is called 
a selfdual `string'.) We consider the case when $\eta^M=({\bf 0},g)$ 
 is aligned along $x^5$ coordinate,
 which we take to be an isometry direction.
That is the soliton (string) is aligned along $x^5$. We consider 
the ansatz
\be\label{hl2} 
X^I(x^m)=\delta^{I8} \phi(x^m),~~~~F_{0m}=\pm g \partial_m\phi .
\ee
This solitonic configuration is a solution of equations \eqn{hl1} provided
\be \phi(x^m)=\phi_0+\sum_i{2 q_i\over|x-x_0^i|^2}
\ee
where fields depend upon all M5 world-volume coordinates $x^m$ ($m=1,2,3,4$)
except $x^5$.
Here $x_0^i, q_i$ are the soliton parameters such as 
positions and charges.
The supersymmetry \eqn{susy2ap} is preserved when 
\be
(1\mp \Gamma^0\Gamma^5\Gamma^8)\epsilon =0
\ee
Since  one of the scalar fields $X^8$ representing a transverse coordinate
 is excited, we  have a 
description in which  M2-brane extended along $x^5$-$x^8$ plane ends 
on  M5-brane. The intersection is  along the common direction $x^5$. 
Such a solitonic configuration will divide M5 world-volume into two 
halves along $x^5$, with a bump on the brane. 
The electric field,  $E_m\equiv F_{0m}$, due to
 string soliton dies off as $1/|x-x_0|^3$, while it is sharply peaked near 
its location at $x_0$. 
\item ii) We also consider a magnetic configuration such as the monopole
in \cite{howe}. We again take $\eta^M=({\bf 0},g)$ aligned 
along $x^5$ as above but
we also consider $x^4$ to be  another isometry direction. The
 remaining spatial coordinates are denoted by
$x^a$ with index $a=1,2,3$. Over this 3-dimensional Euclidean sub-space 
we have a magnetic monopole solution given by
\be
F_{ab}=\mp g \epsilon_{abc} \partial_c \phi, ~~~X^8(x^a)=\phi(x^a)=
\phi_0+\sum_i{2 p_i\over|x-x_0^i|}  
\ee
which solves the equations of motion in \eqn{hl1}. 
For the supersymmetry variations to vanish
we require following condition on the constant spinors
\be
(1\mp \Gamma^0\Gamma^4\Gamma^8)\epsilon =0.
\ee
\ei
Thus the 6D Abelian gauge theory admits ${1\over 2}$-BPS  `string'  and
monopole like solutions, first discussed by \cite{howe} 
in the context of M5-branes.
Also notice that each of these supersymmetric solutions have at least 
one isometry direction. It is also evident from our construction
that there are no stable point-like 
solutions in the theory.

\subsection{Uplifting of 5D super-Yang-Mills to 6D}
From the Abelian exercise we learnt that it is possible to 
construct gauge theories in 6D which may well describe M5-brane. 
We now show that it is also  possible to 
uplift  5D SYM to six dimensions with the help of a space-like 
fixed vector $\eta^M$ in \eqn{vec1}.
We find that a  6-dimensional non-Abelian gauge action  including the 
fermions can be written as
\bea\label{act3a}
S_{M5}(A)&\equiv&\int d^6x \tr \bigg[
-{1\over 12 \eta^4} ( \eta_{[M}F_{NP]})^2  
 -{1\over2} (D_M X^I)^2 +{1\over4}(\eta)^2( [X^I,X^J])^2\br
&&~~+{i\over 2}\bar\Psi\Gamma^MD_M\Psi
-{1\over 2}\eta_M\bar\Psi\Gamma^M\Gamma^I[X^I,\Psi]
  \bigg]
\eea
where the field strength
$F_{MN}=\partial_{[M} A_{N]} -i[A_{M}, A_{N}]$ 
is the Yang-Mills field strength.
The scalar fields $X^I~(I=6,7,...,10)$ are in the adjoint of $SU(N)$.
These correspond to the fact that there are $N$ parallel M5-branes.
The covariant derivatives are
\bea
&& D_M X^I= \partial_M X^I -i[A_M,X^I],~~~ D_M \Psi= \partial_M \Psi 
-i[A_M,\Psi] .
\eea
Interestingly the 5D supersymmetry variations \eqn{susy1} can also be lifted 
to a six-dimensional covariant form
\bea\label{susy2}
&& \delta_s X^I= i\bar\epsilon 
\Gamma^I\Psi,~~~  \delta_s A_{M}= i\bar\epsilon \eta^N\Gamma_{MN}\Psi\br
&& \delta_s \Psi= {1\over3!} {1\over \eta^2}\eta_{[N} F_{MP]}
\Gamma^{NMP}\epsilon 
+D_M X^I\Gamma^M \Gamma^{I}\epsilon
-{i\over 2}[X^I,X^J]\Gamma^{IJ}\eta_M\G^M\epsilon
\eea
under which the action \eqn{act3a} will close on-shell. 
Obviously the theory  possesses 
 a global $SO(5)$ symmetry 
under which five scalars $X^I$ are rotated, which 
is identical to the case of  5D SYM theory. 

We now look at the gauge symmetry possessed by the  
6D gauge action \eqn{act3a}.  The  gauge transformations are
\bea\label{gt1}
&&  A_{M}\to A'_{M}=U^{-1}A_{M} U - i U^{-1}\partial_{M} U\br
&&  X^I\to X'^I=U^{-1} X^I U, ~~~  \Psi\to \Psi'=U^{-1} \Psi U
\eea
under which the action remains invariant. 
Here $U(x)$ is an $SU(N)$ element. 

Notice that the gauge kinetic term in the action \eqn{act3a} 
 has an axial form. Due to the presence of a constant  vector
$\eta^M$, the action is only  covariant but
 not Lorentz invariant.
Any preferential choice of   vector $\eta^M$ 
will break 6D  covariance spontaneously.

\subsection{Compactification to 5D SYM}
We now check explicitly whether we  get  
SYM theory upon  compactification on a circle. 
For simplicity we will set fermions to zero here.
We  also separate the constant 6-vector $\eta^M$ as 
$$\eta^M=(\eta^\mu,\eta^5)$$
where indices $\mu,\nu=0,1,...,4$. 
Without loss of generality let us now assume that
$\eta^M=({\bf 0}, g)$, so that it 
is aligned along 
 $x^5$,  which we are taking
 to be the direction of compactification. 
(We shall also discuss another case where it will not be so.)
It is clear that $A_5$ is an auxiliary field. Consequently the
constraint equation is
\be
D_5X^I=\partial_5X^I-i[A_5,X^I]=0
\ee
So we can easily take all $X^I$'s to be independent of  $x^5$ and
 set $A_5=0$.  We also take 
$A_\m(x^\mu,x^5)=A_\mu(x^\mu)$.
From our ansatz we will have
  $\eta_{[5} F_{\mu\nu]}= \eta_5 F_{\mu\nu}=g F_{\mu\nu}$,
so the 6D action would reduce as
$$-\int d^6x {1\over 12\eta^4} (\eta_{[M} F_{NP]})^2 \to
-\int d^5x {1\over 12\eta^4} 3 (\eta_{[5} F_{\mu\nu]})^2 =
-\int d^5x {1\over 4 g^2} (F_{\mu\nu})^2 $$
To be precise, upon compactification 
there would be a volume factor in 5D theory,
for example radius of compactification $R$, which
 can be absorbed in the definition of 5D YM coupling 
as $R/g^2\equiv 1/g_{_{YM}}^2$ (see also Eq.\eqn{hji1}),     
and in the rescaling of the fields as
$X^I\to \sqrt{R} X^I$,
$\Psi\to \sqrt{R} \Psi$.
From this we determine that the action \eqn{act3a} indeed 
reduces to 5D SYM action \eqn{eq1} while the 
supersymmetry transformations  \eqn{susy2} 
reduce to the transformations given in 
\eqn{susy1}.

We now consider a slightly different situation where 
$\eta^M$ is not be aligned along the isometry direction $x^5$, along
 which we are compactifying.
So we take $\eta^M=(n^\mu,0)$ with $(n^\mu)^2=g^2>0$. Here
 the gauge field components reduce as 
\bea
A_5(x^\mu,x^5)=\phi(x^\mu),~~~~ 
A_\mu(x^\mu,x^5)=A_\mu(x^\mu) 
\eea
So $A_5$  gives rise to a nontrivial scalar $\phi$. 
Taking  all fields $X^I(x^\mu,x^5)$ to be independent of $x^5$, 
 the gauge action \eqn{act3a} 
reduces to the following 5D action (bosonic)
\bea\label{actred2}
S_{5D}
&=&\int d^5x \tr \bigg[
-{1\over 12 n^4} ( n_{[\mu}F_{\nu\lambda]})^2  
-{1\over 4 n^4 } ( n_{[\mu }D_{\nu]}\phi)^2  
 -{1\over2} (D_\mu X^I)^2 \br
&&+{1\over4}n^2( [X^I,X^J])^2
+{1\over2}([\phi,X^J])^2 \bigg]
\eea
Note  the new scalar field $\phi$  also has a potential term of its own. 
This is  an unfamiliar form for  5D Yang-Mills action,  
the difference is that it is written 
with the help of an axial-vector $n^\mu$. We could easily see
that the bosonic content of the theory \eqn{actred2} is the same
as that of standard super-Yang-Mills  \eqn{eq1}. 
Only new thing is that with the help of 
 $n^\mu$ we have been able to pull  $\phi$ out of $A_\mu$, 
 so that only  2 d.o.f. are  
contributed by $A^\mu$. It is again due to the fact that 
 $A_\mu$ kinetic term has axial nature. To see this explicitly, 
we should first take  $n^\mu$ to be
aligned along some spatial direction, say $x^4$, which will reduce 
$A_4$ component to an auxiliary field. 
In the next step we can take $A_4=0$ and  all  fields  to be
 independent of  $x^4$ coordinate.
So that whole dynamics will get confined to  $x_4=0$ hyper-plane. 
Effectively this can be viewed as if 
the 5D theory \eqn{actred2} is  reduced or compactified 
to four dimensions where
it would describe  D3-branes. 
Note that  $\phi$ along with $X^I$'s
constitute  six scalars of 4D SYM.

\section{A local  $\eta^M(x)$ and Lorentz invariance}
In order to promote the 6-dimensional theory to the status of a conformal 
theory with Lorentz invariance it would be essential 
 to lift the  `constant'  vector $\eta^M$ 
to the status of a local field $\eta^M(x)$. That is, 
$\eta^M$ needs to behave like a local Abelian field.  
We can accomplish this task with the help of a
Lagrange multiplier  4-form potential $C_{(4)}$. 
Let us consider the gauge action \eqn{act3a} (bosonic part only), 
especially in the following form
where $\eta^M$ appears  outside the derivatives everywhere 
\bea
S_1(A_M,X^I)\sim\int d^6x \tr \bigg[
-{1\over 12 \eta^4} ( \eta_{[M}F_{NP]})^2  
 -{1\over2} (D_M X^I)^2 +{1\over4}\eta^2([X^I,X^J])^2\bigg] .
\eea
We  now replace $\eta^M$ by a local field $\eta^M(x)$ and 
introduce a Lagrange multiplier term
 \bea\label{ju1}
S=  S_1(\eta^M(x),A_M,X^I) -\int \eta_{(1)}\wedge dC_{(4)}.
\eea 
where  $\eta^M(x)$  appears without  derivatives so it is an auxiliary
field.
As $\eta^M$ is an Abelian field it is expected that
 the action \eqn{ju1} should have an invariance
under gauge transformations like $\eta_M\to \eta_M +\partial_M\zeta$.
\footnote{Such a combination of an auxiliary  $u^M$ vector and  4-form 
potential as a multiplier field has been used in self-dual chiral field
models also \cite{pst96}. 
I am grateful to  M. Tonin for this very helpful communication.}
This local $U(1)$ symmetry can be realized provided
we appropriately  replace $\eta^M(x)$ in the action \eqn{ju1}   
by a  gauge invariant
 combination such as 
$${\hat\eta}^M\equiv (\eta^M-\partial^M a).$$ 
Here $a(x)$  is a new (axionic) scalar  field. 
The complete 6D gauge action then could be written as
\bea\label{ju2}
{\hat S}_{M5}&=&\int d^6x \tr \bigg[
-{1\over 12 {\hat\eta}^4} ( {\hat\eta}_{[M}F_{NP]})^2  
 -{1\over2} (D_M X^I)^2 +{1\over4}{\hat\eta}^2([X^I,X^J])^2\bigg]
 - \int \eta_{(1)}\wedge dC_{(4)}. \br
&\equiv& {\hat S}_1(\hat\eta^M(x),\cdots)  - \int \eta_{(1)}\wedge dC_{(4)}. 
\eea 
 The new local action $\hat S_{M5}$ in \eqn{ju2} 
has  $U(1)$ invariance under
\bea\label{shi3}
\delta\,\eta_M=\partial_M \zeta, ~~~~ 
\delta \, a = \zeta.
\eea
Due to this local shift $\eta^M$ can always eat up  $a(x)$
such that it  will altogether  disappear from the action \eqn{ju2}. 

The $C_{(4)}$ variation of the action \eqn{ju2} provides the
equation of motion 
 \be\label{shi2} 
d\eta_{(1)}=0 .
\ee
Thus on-shell $\eta^M$ will be a constant
vector upto  total derivative term. 
For example, a generic solution of \eqn{shi2}
can be taken as $ \eta_M=g_M+\partial_M\lambda$, where  
$g_M$ being a constant 6-vector. 
 With the help of  shift symmetry \eqn{shi3}
we  fix a gauge where $a(x)=\lambda(x)$, so that the gauge 
invariant quantity $\hat\eta^M$ is a constant, i.e.
\be
{\hat\eta}_M=g_M \ee
However, we should clarify that it is not  guaranteed from here
that the norm 
$({\hat\eta}_M{\hat\eta}^M)$ will always 
be positive definite, it may even be null
or a negative quantity.
But we shall always pick the vacua in which ${\hat\eta}^M$ is 
an spatial vector.

\leftline{\bf  Scaling symmetry:}

The action \eqn{ju2} as usual has an in-built $SU(N)$ gauge symmetry.
We now comment on the scaling symmetry possessed by this action. 
The scaling (mass) dimensions of the  fields are assigned as
\bea
[\eta^M]= -1, ~~[A_M]=1, ~~[X^I]=2,~~[C_{(4)}]=6,
~~[a]=-2
\eea
Note, there is no dimensionful parameter in the action \eqn{ju2}. 
Thus this action
will exhibit an invariance if spacetime coordinates are scaled 
homogeneously and the fields  according to their scaling dimensions.

\leftline{\bf The Vacua:}

Note that the  $N\times N$ constant diagonal matrices 
\be
X^I= {\rm diag}(x^I_{1},x^I_{2},\cdots,x^I_N),~~~~I=6,7,\cdots,10  
\ee
are the simplest solutions of the theory. These give rise to the
moduli space $({\cal R}^5)^N/S^N$ which corresponds to
 $N$ M5-branes placed on a  transverse flat space ${\cal R}^5$.

\leftline{\bf  KK states, 4D instantons and quantized flux:}

The variation of the action ${\hat S}$ with respect to the field $\eta^M$ 
 implies an equation
\bea\label{shi6}
{\delta {\hat S_1}\over \delta \eta_M}={1\over 4!}
 \epsilon^{MNPQRS} \partial_N C_{PQRS}
\eea
which is an important equation which relates $\eta^M$ with  $C_{(4)}$.
It is also an important relation so far as the quantization of flux
and the YM instantons is concerned. Let us consider an static  configuration
 where all $X^I$'s are vanishing, and 
${\hat\eta}^M=g_0\delta^M_5$. We take $x^5$ to be the isometry direction
and compactify it on a circle of radius $R_5$. We will define 
\be\label{hji1}
{R_5\over g_0^2}={1\over g_{_{YM}}^2}.
\ee
Note that, in our convention
 ${R_5\over g_{_{YM}}^2}$ is a dimensionless quantity and it is consistent with
 other expectations \cite{Douglas:2010iu}. 
Consider now the  YM instantons configuration 
$$F_{ij}=\tilde F_{ij}=
{1\over 2!}\epsilon_{ijkl}F^{kl}$$
 living  on the Euclidean space spanned by $x^1,\cdots, x^4$ coordinates. 
Then Eq.\eqn{shi6} implies 
 \bea\label{qua1}
&&{R_5\over 8 g_0^2} \int d^4x \tr F_{ij} F^{ij} 
= R_5g_{0}\int_{\Sigma} dK_{(3)}
\eea
using the relation Eq.\eqn{hji1} we obtain
\bea\label{qua1k}
&&{1\over 8 g_{_{YM}}^2} \int d^4x \tr F_{ij} F^{ij} 
= (R_5)^{3\over2}g_{_{YM}}\int_{\Sigma} dK_{(3)}. \eea
For the above static configuration we have introduced 
a 3-rank Euclidean tensor $K_{ijk}\simeq C_{ijk0}$ and the volume factor $\int dt$ cancels out on 
the both sides of the equation. 
Since the l.h.s. of \eqn{qua1k} counts the instanton number so it would be
 quantized
\bea\label{qua2k}
{1\over 8 g_{_{YM}}^2} \int d^4x \tr F_{ij} F^{ij}={4\pi^2 k\over g_{_{YM}}^2} 
\eea
where index $k\in Z$ counts the instantons.\footnote{ In order to make a
better  link with \cite{Douglas:2010iu,lps} works, one may instead choose 
a different length parameter  $R_5'=g_{YM}^2/(4\pi^2)$ on the r.h.s.
of \eqn{qua2k}.   $R_5'$ can be identified with the radius of the circle 
in the  6D CFT compactification \cite{lps}.
 But we have not worried much about that here as our claim  
is to establish a  quantized flux. }

  We then obtain from \eqn{qua1k} 
\be\label{qua1}
 {1\over 4\pi^2} \int_{\partial\Sigma} K_{(3)}=
{k\over (R_5g_{_{YM}}^2)^{3\over2} }\ .
\ee
The equation \eqn{qua1} implies an existence of a quantized 4-form flux 
threading  
the Euclidean 4-fold $\Sigma$, having a boundary $\partial \Sigma$. 
The 4D Yang-Mills instanton number $k$  has an interpretation as the
 momentum $p_5=k/R_5'$ carried by the KK states of 5D SYM
\cite{Douglas:2010iu, lps}. Accordingly  
these KK states need to be taken into account in 5D SYM if the 
instanton number is nonvanishing. We have shown that the  YM instantons 
for the 6D gauge theory compactified on $S^1$, does
imply a nonvanishing quantized flux.

\section{Conclusions}
We have  lifted 5D SYM  theory to six dimensions
keeping the  non-Abelian  
structure intact. 
The whole procedure can be made 
Lorentz covariant provided we assume the existence of a  
space-like  `coupling constant'  6-vector $\eta^M$,
such that $\eta^M\eta_M=g^2 >0$.   
The norm of  $\eta^M$  manifests as the super-Yang-Mills coupling 
constant  in five dimensions. The theory can also be made
fully localized. The 
Lorentz invariance is regained with the help of a
Lagrange multiplier 4-form field. 
It should nevertheless be explored further if our procedure leads to 
desired properties like conformal symmetry for these 6D gauge theories. 
We have shown that 
when we treat $\eta^M$ as an auxiliary  field
the theory becomes  scale invariant. 
We also find that the 6D gauge theory does admit 1/2-BPS  `string' and 
monopole like solutions, first obtained by \cite{howe} in the context 
of M5-branes. From our construction we find 
that there are no stable point-like 
solutions in our theory.

\vskip.5cm
\noindent{\it Acknowledgement:}\\
This work got initiated during the workshop  "Indian String Meeting"
 ISM'2011, at Puri and the conference  "New Trends in Field Theories", at
BHU, Varanasi. I take this opportunity  to  thank the Conveners 
of both these  meetings.


\appendix{
\section{Yang-Mills theory in axial gauge}
Let us take the case of pure Yang-Mills theory in  $5$ dimensions.
The standard Yang-Mills action in axial gauge is written as
\bea\label{app1}
S_5=\int d^{5}x ( -{1\over 4 g^2}\tr (F_{\mu\nu})^2  -{\alpha\over 2} 
\tr (\eta^\mu A_\mu)^2)
\eea
where indices $\mu=0,1,...,4$ and $\eta^\mu$ is a fixed 5-vector. 
 The condition
$$\eta^\mu A_\mu=0$$ is known as the axial gauge.
The term  $(\eta^\mu A_\mu)^2$ in the above action explicitly  breaks 
the gauge invariance under the transformations  
\bea\label{gt2}
&&  A_{\m}\to A'_{\m}=U^{-1}A_{\m} U - i U^{-1}\partial_{\m} U
\eea
The gauge transformations \eqn{gt2} have to be such that 
 \be
\eta.A=\eta.A' 
\ee
which puts following restriction on gauge functions 
\be
\eta^\m U^{-1} \partial_\m U=0 ,
\ee
One however could choose a particular local frame in which 
$\eta^\mu=(0,\cdots,0,g)$  
so that $A_4=0$, in that case $U$ has to be independent of  $x^4$.

A distinct but less familiar Yang-Mills action  involving a fixed vector 
can be written as
\bea\label{app2}
S=-\int d^5x {1\over 12 (\eta)^4}\tr ( \eta_{[\mu}F_{\nu\lambda]})^2  
\eea
where  $\eta^\mu\eta_\mu > 0$. That is, vector $\eta^\mu$ 
is space-like.
The important thing about  action \eqn{app2} is that it
 has explicit invariance under the gauge transformations \eqn{gt2}
where $U$ is unrestricted. 
Now if we locally take $\eta^\mu$ to be aligned along 
 $x^4$, i.e. $\eta^\mu=({\bf 0},g)$, the
$A_4$ component of the gauge field becomes auxiliary.
An special case is   $A_4=0$ and take $A_\alpha$'s 
independent of  $x^4$ coordinate.
Then above gauge action \eqn{app2} immediately
 reduces to the 4D Yang-Mills action
\bea\label{app2a}
S=-\int d^4x {1\over 4  g^2} \tr( F_{\alpha\beta})^2  
\eea
where $\alpha,\beta =0,1,2,3$.
The special advantage with the action \eqn{app2} is that it has
explicit gauge invariance. While
it  contributes one less degree of freedom compared to the 
usual YM field. 
Thus the  situation is completely different 
from that of the YM action in \eqn{app1}.
 
Let us comment on what will be the situation if $\eta^\mu$ is 
taken to be a time-like vector, $\eta^2=-g^2$. We  consider
the case $\eta^\mu=(g,0,0,0,0)$. One can see that 
the action \eqn{app2} 
 reduces to an Euclidean  action in 4D
\bea\label{app2b}
S_E=\int d^4x{_{E}} {1\over 4  g^2} \tr ( F_{ij})^2 . 
\eea

\section{Is there a tensor version of the 6D gauge theory?}
We may wonder if 6D gauge theory  could also be described 
in terms of tensors. For this we would need to introduce  
a tensor field $B_{MN}$ in the adjoint representation 
of the  $SU(N)$ gauge group. 
It is challenging how to first define a gauge invariant
non-Abelian tensor field strength involving
tensor fields such that it includes {\it self-interactions},   like the
YM gauge fields do. Thus
with the help of constant vector $\eta^M$ at our disposal 
we define a non-Abelian tensor field strength
as
\bea\label{jk2}
H_{MNP}
&=&\partial_{[M} B_{NP]}-{i\over 2}\eta^Q([B_{QM}, B_{NP}]
+~{\rm cycl.~perm.~of~}~_{M,N,P}) 
\eea
The main feature here is that there is a $[B,B]$ commutator as we encounter
in the Yang-Mills case.
But we can also introduce vector fields through the relation 
\be\label{def2}
  \eta^N B_{NM} \equiv A_M
\ee
This way of introducing a vector fields is distinct in the sense
 that we will immediately have an axial gauge condition
\bea\label{def1}
&& (i)~ \eta^M A_M=0 , 
\eea
and also then we can  write
\bea\label{def2a}
&& (ii)~B_{MN}\equiv {1\over ~(\eta)^2}\eta_{[M}A_{N]}. 
\eea
Using \eqn{def2} to \eqn{def2a} we find 
\bea\label{jk1}  
H_{MNP}
&\equiv &\partial_{[M} B_{NP]}-{i\over 2}([A_{M}, B_{NP}] 
+~{\rm cycl.~perm.})\br  
& \equiv& {1\over (\eta)^2}\eta_{[N} F_{MP]}
\eea
where $F_{MN}$ is usual YM field strength.
Thus our definition \eqn{def2} is analogous to including an 
axial gauge condition for  the Yang-Mills fields $A_M$ in 6D. 
That is  in describing  tensor field through a gauge field in
\eqn{def2} we would be in the axial gauge set up. 
We should keep in mind that writing the tensor field as in \eqn{def2a}  
means that it is not a
fundamental 2-rank tensor field instead it
is a kind of composite tensor and thus carries 
only partial degrees of freedom 
constituted primarily by the YM fields.
\footnote{In alternative formulations like (2,0) tensor CFTs, 
one considers the self-dual tensor field strengths,
$H_3=\star H_3$, 
 in order to  halve the physical degrees of freedom
carried by a tensor field. While, in our formulation 
here we are having a composite nature of the tensor field 
avoiding or circumventing the self-duality criterion altogether.} 

The  equation \eqn{jk2} suggests that the tensor field 
is indeed self-interacting but the last equality in \eqn{jk1} also
suggests that, although we have defined a 3-form field strength in reality 
we are effectively dealing with Yang-Mills fields only  in disguise. 
The 6D tensorial form of the action \eqn{act3a} can be written as
\bea\label{act3a0}
S_{M5}(B)&=&\int d^6x \tr \bigg[
-{1\over 2.3!} ( H_{MNP})^2  
 -{1\over2} (D_M X^I)^2 +{1\over4}(\eta_M [X^I,X^J])^2\br
&&~~+{i\over 2}\bar\Psi\Gamma^MD_M\Psi
-{1\over 2}\eta_M\bar\Psi\Gamma^M\Gamma^I[X^I,\Psi]
  \bigg],
\eea
The covariant derivatives can be defined as
\bea
&& D_M X^I= \partial_M X^I -i\eta^N[B_{NM},X^I],~~~ D_M \Psi= \partial_M \Psi 
-i\eta^N[B_{NM},\Psi]
\eea
While the susy transformations can be written as
\bea\label{susy2i}
&& \delta_s X^I= i\bar\epsilon \Gamma^I\Psi\br
&& \delta_s B_{MN}= i\bar\epsilon \Gamma_{NM}\Psi\br
&& \delta_s \Psi= {1\over3!}  H_{MNP} \Gamma^{MNP}\epsilon 
+D_M X^I\Gamma^M \Gamma^{I}\epsilon
-{i\over 2}[X^I,X^J]\Gamma^{IJ}\eta_M\G^M\epsilon
\eea
under which the action \eqn{act3a0} closes on-shell. 

The gauge transformations of the 
$B$-fields can be written as
\be
 B_{MN}\to B'_{MN}=U^{-1}B_{MN} U - i g^{-2}U^{-1}\eta_{[M}\partial_{N]}U .
\ee
while  rest of the fields transform in the same way as in \eqn{gt1}.
This indicates that $B_{MN}$ is not the fundamental 2-rank field. 
As discussed above these tensors
contributes only 3 physical degrees of freedom. 
}

\end{document}